\RequirePackage{fix-cm}
\documentclass[twocolumn]{svjour3}          
\smartqed  
\usepackage{amsmath}
\usepackage{nccmath}
\usepackage{amsfonts}
\usepackage{graphicx}
\usepackage[dvipsnames]{xcolor}
\usepackage[numbers,sort,compress]{natbib}
\begin{document}

\title{Data-driven nonlinear modal identification of nonlinear dynamical systems with physics-constrained Normalizing Flows
}


\author{Abdolvahhab Rostamijavanani \and Shanwu Li         \and
        Yongchao Yang 
}


\institute{Corresponding Author: Abdolvahhab Rostamijavanani \at
              Department of Mechanical Engineering - Engineering Mechanics, Michigan Technological University, Houghton, Michigan 49931, USA\\
              \email{arostami@mtu.edu}               
}


\maketitle

\begin{abstract}
Identifying the intrinsic coordinates or modes of the dynamical systems is essential to understand, analyze, and characterize the underlying dynamical behaviors of complex systems. For nonlinear dynamical systems, this presents a critical challenge as the linear modal transformation, which is universal for linear systems, does not apply to nonlinear dynamical systems. As natural extensions to linear normal modes, the nonlinear normal modes~(NNMs) framework provides a comprehensive representation of nonlinear dynamics. Theoretically, NNMs may either be computed numerically or analytically from the closed-form models or equations of dynamical systems, or experimentally identified from controllable input-output tests, both of which, however, are typically unknown or unavailable practically. In this study, we present a physics-integrated Normalizing Flows deep learning-based data-driven approach which identifies the NNMs and the nonlinear modal transformation function of NNMs using measured response data only. \textcolor{black}{Specifically, we leverage the unique features of the Normalizing Flows model: 1)   the \textit{independent} latent spaces, naturally spanned by the Normalizing Flows, are exploited to facilitate nonlinear modal decomposition; 2) the invertible transformation through the Normalizing Flows, enabling efficient and accurate nonlinear  transformation between original and modal coordinates transformation. Therefore, our framework leverages the independency feature and invertibility of Normalizing Flows to create a model that captures the dynamics of unknown nonlinear dynamical systems. This enables the identification of nonlinear normal modes through data-driven methods, while also preserving the physical interpretability and generalizability of resulting invariant manifolds and long-term future-state predictions for a wide range of physical systems.} For method validation, we conduct numerical experiments on multi-degree-of-freedom (MDOF) Duffing systems and velocity fields of flow passing a cylinder in the laminar regime. We present the performance of the presented method in identifying the nonlinear manifolds of a dynamical system under different energy levels, and compare the presented method with the Proper Orthogonal Decomposition~(POD) method. It is observed that the identified NNMs achieve higher representation accuracy than the POD method using the same dimension of intrinsic coordinates or modes. \textcolor{black}{We also discuss the limitation of the 
 presented framework on high-dimensional dynamical systems, where a dimension reduction scheme is applied in the flow field case study.}

\end{abstract}

\section{Introduction}\label{sec:1}


Complexity in a system is commonly characterized by nonlinear dynamical behaviors, whose characterization remains a perennial challenge in a wide range of fields including engineering fields~\cite{Strogatz1994} \textcolor{black}{such as modal analysis and system identification}. \textcolor{black}{For modal identification and characterization,} in contrast to linear systems that can be modelled using linear modal transformations as superpositions of linear normal or eigen modes~\cite{Heylen1997}, which provide an exact description of their underlying linear dynamical characteristics and facilitate linear reduced-order modeling~\cite{Haller2016,Komatsu1991,Likins1967,Touze2008,Bladh2002,BladhR.2001,Peeters2009,Vakakis1997,Kerschen2009}, nonlinear dynamical systems can not be represented precisely by such a general linear framework. As an example, linear modal analysis methods such as proper orthogonal decomposition (POD), independent component analysis (ICA)~\cite{stone2004independent,kerschen2007physical}, and dynamic mode decomposition (DMD)~\cite{kutz2016dynamic,schmid2010dynamic,schmid2011applications} are not applicable for highly nonlinear dynamical systems because significant error occurs when using these linear methods. \textcolor{black}{Even though ICA shows reasonable performance for linear structural dynamics, it fails when there is high damping in systems~\cite{stone2004independent,kerschen2007physical}. }\textcolor{black}{Koopman Mode Decomposition~(Koopman operator)~\cite{Mezic2005,Mezic2013} describes nonlinear dynamical systems by linearly projecting observables onto Koopman eigenfunctions~\cite{Koopman1931}. However, obtaining an observable function that can transform nonlinear dynamics into a new state space, where the underlying nonlinear dynamics can be approximated linearly, is a difficult task as it requires infinite dimensions}. Thus, discovering a nonlinear generalization of modal transformation is a crucial problem in nonlinear structural dynamics and fluid dynamics.

~\textcolor{black}{Nonlinear normal modes (NNMs), as a nonlinear modal analysis technique first introduced by Rosenberg~\cite{Rosenberg1960} and extensively studied by many researchers~\cite{kerschen2009nonlinear,vakakis1991analysis,rand1971higher,shaw1991non,shaw1993normal}, are natural extensions of linear normal modes~(LNMs). Like LNMs, NNMs are also intended to capture the intrinsic invariance properties of nonlinear dynamics.} Fundamental work by Shaw and Pierre~\cite{Shaw1993} provided an extension of the concept of LNMs for nonlinear systems by mapping the physical coordinates to the nonlinear modal coordinates through a nonlinear transformation. These studies provide insights into a general and interpretable modal transformation for nonlinear systems. The majority of research on identifying defined nonlinear modal transformations has sought to numerically compute the closed-form model of the system~\cite{Kuether2014,Peeters2009a,Ponsioen2018}. Nevertheless, the closed-form equations are mostly unknown or unavailable for real-world systems, only observations and measurements available. Also, using Taylor series expansions to identify modal coordinates is merely an approximation of nonlinear normal modes analysis \cite{shaw1991non}. \textcolor{black}{Hence, data-driven methods are well-suited for addressing the aforementioned limitations.}

  Data-driven modeling of nonlinear systems has been facilitated by leveraging machine learning and deep learning techniques. \textcolor{black}{Deep learning's fundamental architecture, known as deep neural networks (DNNs), offers remarkable modeling capacity and learning flexibility to capture the intrinsic features of complex systems in a hierarchical manner. The universal approximation theorem~\cite{Hornik1990,Akhtar2018}, states that a DNN with an adequate number of neural units and nonlinear activations can represent intricate functions, such as nonlinear modal transformation functions and temporal evolution functions of dynamics. DNNs also allow for the adaptive design of network architecture for various tasks, including the identification of NNMs and dynamics.} There are several deep neural networks that have been introduced by researchers to model well-known representations of dynamical systems, which include Koopman operators and NNMs methods~\cite{dervilis2019nonlinear,Worden2017,Yeung2019,lusch2018deep,takeishi2017learning}. For training the models, the corresponding architectures require multiple cost functions. Notably, a deep neural network (DNN) framework with axillary networks for Koopman operators was developed to model continuous spectra of nonlinear dynamical systems ~\cite{lusch2018deep}. By embedding new correlation loss functions,nonlinear normal modes and their transformations were identified using deep autoencoders \cite{li2021data}. \textcolor{black}{Typically, adjusting several loss functions is necessary and considerably challenging to achieve a nonlinear transformation between the original coordinates and modal coordinates, as well as the ability to move back and forth between these coordinates and the independence of modal coordinates. Normalizing Flows, an emerging generative model, offer a new efficient and effective deep learning alternative to address these aforementioned challenges; it is explored in this study for data-driven identification of nonlinear normal modes.} 

\begin{figure*}[htbp]
	\centering
	\includegraphics[width=0.9\textwidth]{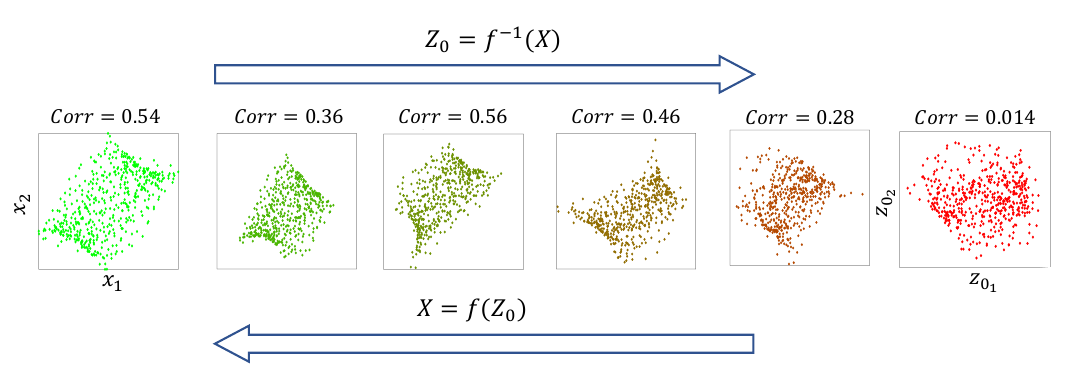}
	\caption{Normalizing Flows concept: Reducing the dependency between two original coordinates of a 2-DOF Duffing system by passing through Normalizing Flows layers. $Z_0$ denotes the decomposed modal coordinates while $X$ corresponds to the original coordinates.}
	\label{FIG:2}
\end{figure*}

\begin{figure}[htbp]
	\centering
	\includegraphics[width=0.5\textwidth]{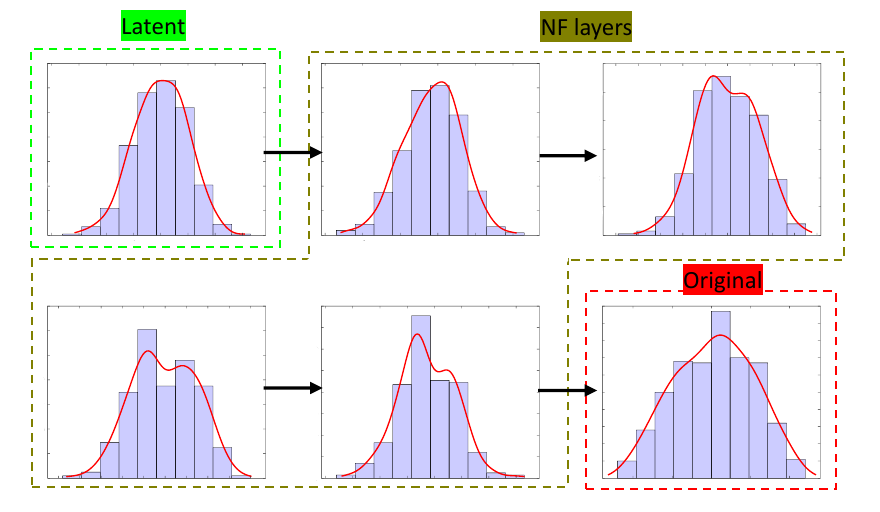}
	\caption{Probability distribution transformation of a dimension of a Duffing system with Normalizing Flows model. Simple PDF in the latent can transfer to more complicated in the original space~(learning the original PDF by transforming a simple PDF with bijectors)  }
	\label{fig_probab}
\end{figure}

Normalizing Flows~(NF) (Fig. \ref{FIG:1}) is an advanced generative model~\cite{urain2020imitationflow,kobyzev2020normalizing,khader2021learning,sun2022probabilistic} that may \textcolor{black}{be used for nonlinear modal analysis (e.g. NNMs) by virtue of its unique specifications}. \textcolor{black}{The salient feature of NNMs, invariance, may be characterized by Normalizing Flows' \textit{independence} feature of the modal coordinates and the associated nonlinear modal transformation.} ~\cite{dinh2014nice,dinh2016density}~(Fig.~\ref{FIG:2}). Such a capability is attributed to theory of Normalizing Flows, whose model tries to estimate the real probability distribution of a variable by transforming a random distribution like Gaussian distribution~(Fig.~\ref{FIG:1} and Fig.~\ref{fig_probab}) into independent distribution. Therefore, it is feasible to perform modal decomposition naturally with Normalizing Flows models. The other unique feature of Normalizing Flows is that the transformation is \textit{invertible}~\cite{papamakarios2021normalizing}. \textcolor{black}{The explicit invertibility within the layers of Normalizing Flows allows us to easily transition between original coordinates and modal coordinates using this feature.}

In this study, exploiting Normalizing Flows models, we present a data-driven deep neural network approach to address the challenges in nonlinear modal identification. Specifically, the lack of closed-form solutions in the modal analysis domain is addressed by data-driven approaches, and unavoidable errors that can occur in \textcolor{black}{coordinates transformation} is rectified by leveraging the invertibility feature of the Normalizing Flows. Additionally, the independent latent spaces, naturally spanned by the Normalizing Flows, are exploited to facilitate nonlinear modal decomposition. Moreover, we incorporate a \textit{dynamics block} into the Normalizing Flows model in order to simultaneously capture the underlying modal dynamics of the studied systems \textcolor{black}{and facilitate long-term future-state prediction}. Therefore, our presented NF-based DNN is capable of not only performing nonlinear modal decomposition, but also predicting the behavior of dynamical systems for a specific range of time.

\section{Methodology: Normalizing Flows concept}
As a generative model, Normalizing Flows produces tractable distributions, which make density estimation efficient and precise. Other generative models such as generative adversarial networks (GANs) and variational
auto-encoders (VAEs) do not explicitly learn the probability function of training data. GANs produce similar data to fool the discriminator in a min-max game and reach a saturation point when the discriminator cannot distinguish the fake samples~\cite{goodfellow2020generative}. With VAEs, the network learns how to identify variational inference in latent space and to produce data using a decoder~\cite{kingma2013auto}. However, neither GANs nor VAEs are capable of learning the real probability density functions (PDFs) of real data. 

Normalizing Flows model is a rigorous generative method that learns the real PDF of a dataset by using some invertible and differentiable functions. Normalizing Flows involves creating a random variable $\mathcal{X}$ with a complicated distribution $\mathcal{P}$ by applying a function $\textit{f}$ that is both invertible and differentiable to a random variable $\mathcal{Z}$ with a simple distribution. As an example, a standard normal distribution $\mathcal{Z} \sim N(0, 1)$ can be inverted to a target PDF using the following formula:
$\mathcal{ X} = f(\mathcal{Z}) \sim \mathcal{P}$ , $\mathcal{Z}=f^{-1}(\mathcal{X})$. The transferred distribution $\mathcal{P}$ can be calculated by using change of variables:
\begin{equation}
    log \hspace{0.1cm}\mathcal{P}(\mathcal{X})= log \hspace{0.1cm}\mathcal{P}(\mathcal{Z})-log\hspace{0.1cm}|\frac{\partial f}{\partial \mathcal{Z}}(\mathcal{Z})|
\end{equation}
It is, however, not trivial to find a single function (bijector) that transfers the distribution in the desired manner. When the target distribution $\mathcal{P}$ is very complex, a simple $\textit{f}$ (such as a scale or shift function) is not sufficient. The following example illustrates how to create a more complicated PDF by composing bijectors with one another and creating a more complicated chain of bijectors:
\begin{equation}\label{eq2}
    \textit{f}=\textit{f}_{k}\hspace{0.1cm} \circ \hspace{0.1cm}f_{k-1}\hspace{0.1cm} \circ\hspace{0.1cm}...f_{1}
\end{equation}
A Normalizing Flows refers to the transformation of a base distribution (e.g., standard normal distribution) into the more complex target distribution through a series of bijectors after each other:

\begin{equation}
   \mathcal{Z}_{k}=f_{k}(\mathcal{Z}_{k-1})
\end{equation}
Additionally, one can calculate the transformed~(target) distribution by summing the contributions from each bijector:
\begin{equation}
  log \hspace{0.1cm}\mathcal{P}(\mathcal{X})= log \hspace{0.1cm}\mathcal{P}(\mathcal{Z})-\sum_{i=1}^{i=k}log\hspace{0.1cm}|\frac{\partial f_{i}}{\partial \mathcal{Z}_{i-1}}(\mathcal{Z}_{i-1})|
\end{equation}
To achieve the target distribution, one can give each $\textit{f}_i$ some simple functions, such as a scale and shift, followed by a simple nonlinearity, such as a sigmoid or ReLU function. It should be noted that each $\textit{f}_i$ has some parameters (such as scale and shift values), which can be learned from some training data using \textit{maximum likelihood estimation}.
\subsection{Models with Normalizing Flows}
The properties of each Normalizing Flows model are the same: they are invertible and differentiable. By stacking a sequence of bijectors with easy Jacobian determinant computations and easy invertibility, the RealNVP model (Real-valued Non-Volume Preserving)~\cite{dinh2016density}  can be modelled. Non-linear Independent Component Estimation (NICE)~\cite{dinh2014nice} is an alternative to RealNVP that was introduced earlier. Autoregressive models are a type of Normalizing Flows in which the Jacobean matrices can be computed quickly since each $\textit{f}_i$~(equation. \eqref{eq2}) depends on $\mathcal{Z}_{1},...,\mathcal{Z}_i$. Hence, ${\frac{\partial f_{i}}{\partial \mathcal{Z}_{j}}=0}$ whenever $\mathrm{j>i}$ and a lower triangular Jacobean matrix is achieved and the determinant of this matrix is a simple product of the diagonal elements. Also, the joint density $\mathcal{P}(\mathcal{X})$ can be modelled as the product of conditionals $\prod_{i} \mathcal{P}(\mathcal{X}_{i}|\mathcal{X}_{1:i-1})$. Throughout this paper, a masked autoregressive flow model is used.
\begin{figure*}[htbp]
	\centering
	\includegraphics[width=0.95\textwidth]{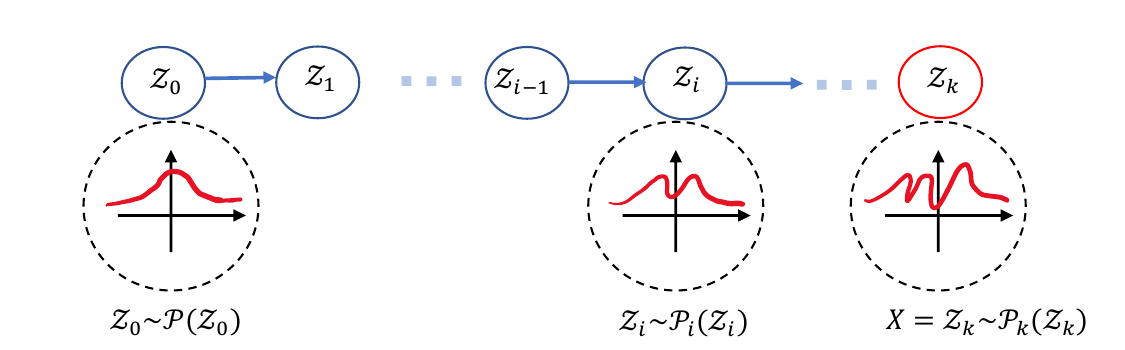}
	\caption{Normalizing Flows concept: Transforming a simple probability distribution function~($\textit{P}_0$) to the original one~($P_k$) which is more complex by using bijectors.}
	\label{FIG:1}
\end{figure*}
\begin{figure*}[htbp]
	\centering
	\includegraphics[width=0.9\textwidth]{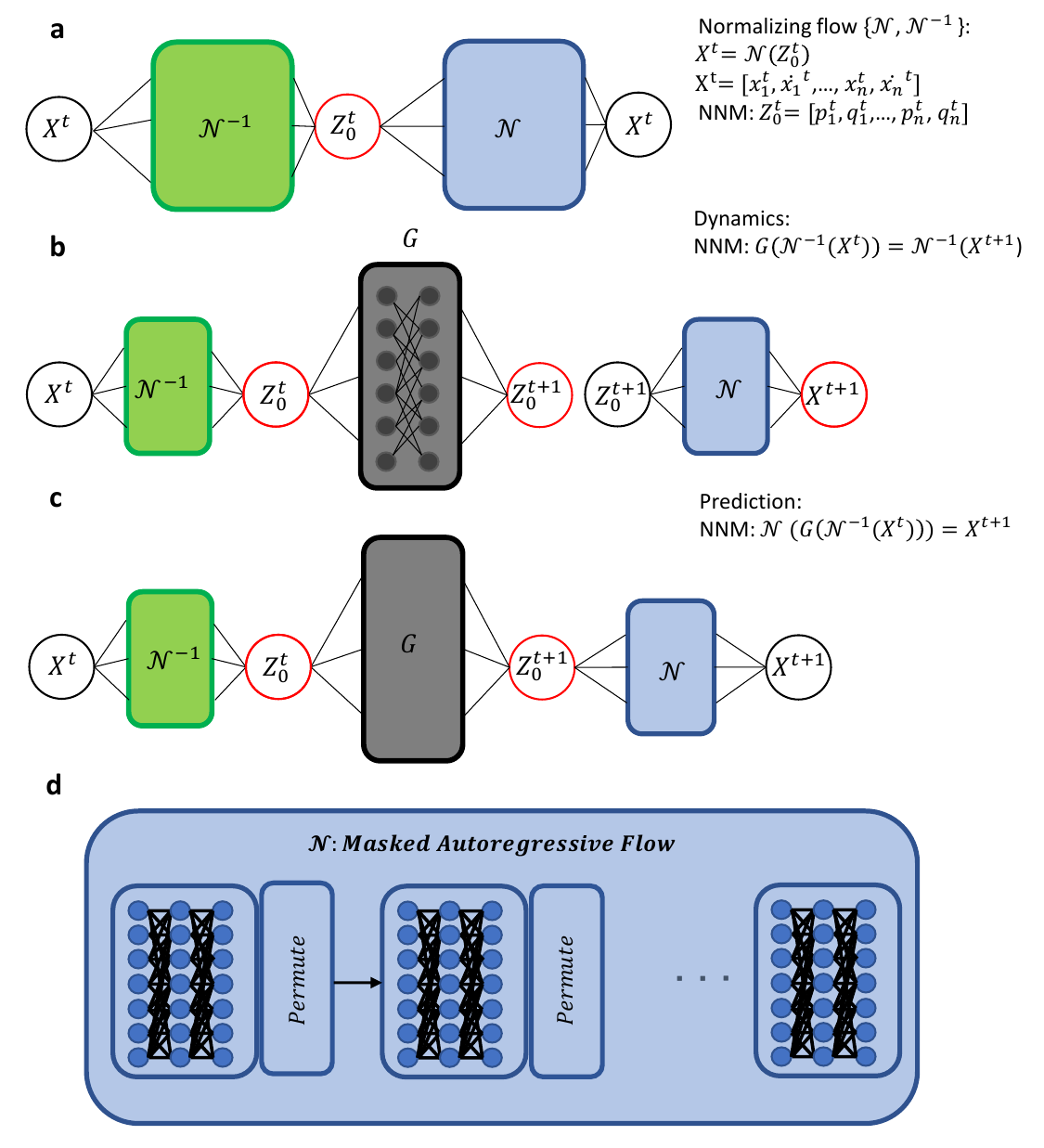}
	\caption{Architecture of the presented physics-constrained Normalizing Flows deep neural network~(NF-DNN). \textbf{(a)} The overall framework consists of a NF model that transfers states~$X^t=(x,\dot x)$ of a system into intrinsic coordinates~$Z_0^t=(p,q)$ using ${{Z_0^t}} = \mathcal{N}^{-1} \left( {{X^t}} \right)$ and then transforms them back to original coordinates by simply inverting the forward process ${{X^t}} = {\mathcal{N} }\left( {{Z_0^t}} \right)$. The subscript $0$ represents the latent space where the base distribution is assumed to be a Gaussian distribution. There are several additional physics-based constraints that can be applied to the intrinsic coordinates ${{Z_0^t}}$ to enforce them to be translated to desired modal coordinates. \textbf{(b)} A dynamics block ($G$) is implemented, which advances intrinsic coordinates forward in time and enforces the equivalence between transferring the next original coordinates and advancing current intrinsic coordinates forward. This ensures the dynamics of the system remain in the identified intrinsic coordinates. \textbf{(c)} By combining Normalizing Flows model and dynamics block, intrinsic coordinates are determined for enabling future-state prediction. Unlike autoencoders where decoders are not exactly the inverse function of encoders, NF is able to perform forward and inverse process without any approximation errors since the process is a direct mathematical inversion. \textbf{(d)} The presented NF model is a Masked Autoregressive Flow model that consists of main layers and Permute layers.}
	\label{FIG:3}
\end{figure*}
\begin{figure*}[htbp]
	\centering
	\includegraphics[width=0.9\textwidth]{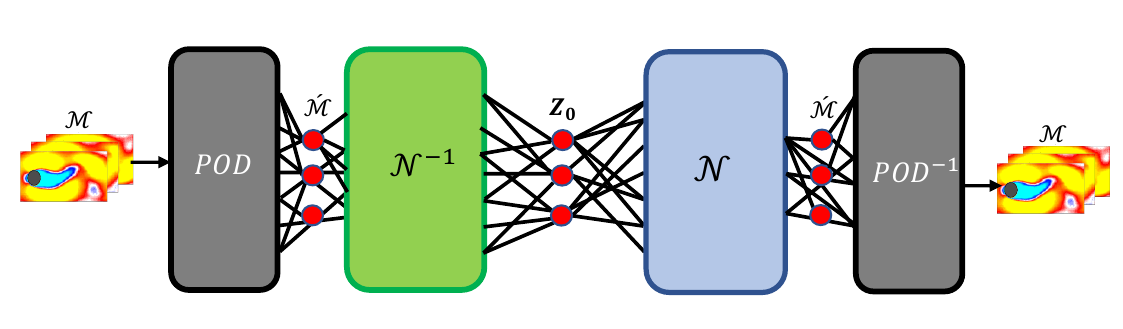}
	\caption{Overall process of mode decomposition for flow fields~($\mathcal{M}$): First a reduced order of flow is achieved by POD transformation and then these modal coordinates~($\mathcal{M}^\prime $) pass through NF layers to be more independent using the natural feature of NF - independence in latent space~($Z_0$). At the end the latent spaces are transferred back to the original flow fields~($\mathcal{M}$) using the inverse function of POD algorithm.  }
	\label{FIG:4}
\end{figure*}
\begin{figure}[htbp]
	\centering
	\includegraphics[width=0.5\textwidth]{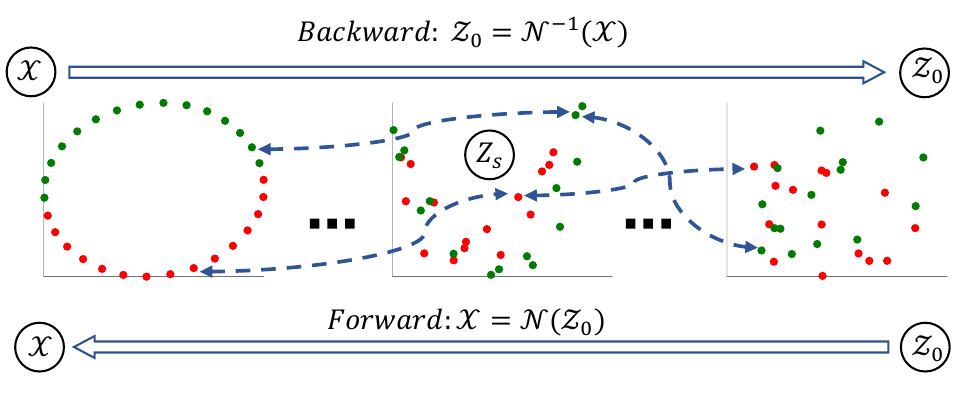}
	\caption{General representation of one-to-one mapping feature of Normalizing Flows models. The size of data retain the same in the latent~(right plot: $Z_0$) and any layer between original and latent (middle plot: $Z_s$) as original space~(left plot: $X$). An instance of Normalizing flow can demonstrate how data with circular dependencies can be converted to an independent state represented by $Z_0$. }
	\label{fig-one-to-one}
\end{figure}
\section{Problem formulation}\label{sec:2}
\subsection{Multi-degree-of-freedom (MDOF) systems}
Consider the free response of an $N$-degree-of-freedom ($N$-DOF) nonlinear system with the general equation of motion:
\begin{equation}\label{eq1}
{\mathrm{M}\ddot x} + {\mathrm{C}\dot x} + {\mathrm{K}x} + {\mathrm{g}}\left( {{{\ddot x}},{{\dot x}},{{x}}} \right) = 0
\end{equation}
\noindent where $\mathrm{M}$, $\mathrm{C}$, and $\mathrm{K}$ are mass, damping, and stiffness matrices, respectively. $ x$ is displacement vector~{(${x\in R^n}$)} and $\mathrm{g}$ is a nonlinear term.
Transforming to the discrete-time state space:
\begin{equation}
\begin{split}
    { X=\{ x,\dot{ x}\}}\\
    { X^{t+1}=\mathrm{F}\left( X^{t}\right)}
    \end{split}
\end{equation}
where ${X} \in R^{2n}$ denotes state space vectors which are measured by sensors or are computed numerically, and $\mathrm{F}$ is the dynamics function which maps current states to states forward in time. \par
NNMs represents the nonlinear dynamics through nonlinear transformations of its intrinsic modal coordinates. As natural extensions of LNMs, NNMs can represent nonlinear dynamical systems with the same number (dimension) of modal coordinates as the original coordinates:
\begin{equation}
    {Z_0}^{t+1}=G\left(Z_0^{t}\right)
\end{equation}
where for second-order ODEs, each NNM modal coordinate, $Z_0$, consists of displacement ($\mathrm{p}$) and velocity ($\mathrm{q}$) fields~(Fig. \ref{FIG:3}), and $\textit{G}$ is the nonlinear transformation function which represents modal state transition. Note that the intrinsic modal coordinates are denoted as $\textit{Z}_0$ which are identified as the NNMs when integrating their physics constraints with the presented deep learning-based data-driven system identification framework.

The aim of using normalizing flows in this context is to convert the original coordinates~($X$) of MDOF systems into modal (NNM) coordinates ($Z_0=\mathcal{N}^{-1}(X)$) and vice versa ($X=\mathcal{N}(Z_0)$), leveraging the \textit{independent} property of NF. This involves using the inverse capability of NF, which transfers a simple distribution, representing modal coordinates, to a more intricate distribution in the original coordinate space (multi-component system response or vibration). By integrating physics constraints related to NNMs, we can establish the modal coordinates as a representation of NNMs.
\subsection{Flow fields}
By combining POD with normalizing flows, the reduced set of modal coordinates obtained from POD is used as input to Normalizing Flows. Then, the probability distribution of the modal coordinates obtained from POD is modelled and their independence is enhanced by the Normalizing Flows. This will enable a more accurate representation of the non-Gaussian and complex distributions that are commonly encountered in fluid dynamics problems. In other words, we seek a nonlinear ('deep') version of POD modal coordinates when dealing with flow fields. To decompose POD modal coordinates~(Fig. \ref{FIG:9} and Fig. \ref{FIG:10}) we discover a function that transforms the original coordinates to a new space where the obtained modal coordinates are independent:
\begin{equation}
    {\mathcal{M}^\prime=\mathcal{N}(Z_0)}
\end{equation}
where $\mathcal{M}^\prime$ denotes the POD modal coordinates of the flow field and ${Z_0}$ is the decoupled version of POD modal coordinates obtained via Normalizing Flows~($\mathcal{N}$). It should be noted that, the focus in case studies involving flow fields is on nonlinear mode decomposition rather than prediction and dynamics.
\section{Normalizing Flows-based deep learning framework to identify NNMs operator}
\subsection{Objective}
There are challenges associated with obtaining accurate coordinates for  NNMs. Using Taylor series expansions to identify modal coordinates is limited to an approximation of nonlinear normal modes analysis \cite{shaw1991non}. Also, most of real-world dynamical systems do not have closed-form solutions and we have only some sensors measurements. Therefore, we develop a data-driven Normalizing Flows framework as shown in Fig. \ref{FIG:3} and the objective of our framework is to identify NNMs operators in order to overcome the challenges mentioned above related to lack of closed-form solutions of dynamical systems and identifying nonlinear modal coordinates of NNMs. Moreover, we aim to obtain a nonlinear version of POD modal coordinates to represent the most essential characteristics of flow fields on an independent basis.

\subsection{Normalizing Flows framework}
\subsubsection{2-DOF Duffing oscillator}
In this session we present an example of a 2-DOF Duffing oscillator as one of case studies. Since this dynamical system is a second-order ordinary differential equation (second ODE), each pair of latent subspace corresponds to the displacement and velocity of one degree of freedom in the Normalizing Flows-based deep neural network (NF-DNN) presented for NNMs. The number of latent dimensions is the same as the original dimension. Therefore, each pair of latent coordinates represents a single nonlinear normal modal coordinate.\par
The presented NNMs-physics-constrained Normalizing Flows (NNMs-NF-DNN) integrates the physics of NNMs into the deep learning. The overall loss function is as following:
\begin{multline}\label{eq 9}
    \mathrm{\mathcal{L}_{NNM}= {\alpha _{\mathcal{N}}}{\mathcal{L}_{\mathcal{N}}} + {\alpha _{corr}}{\mathcal{L}_{corr}} + {\alpha _{evol}}{\mathcal{L}_{evol}}} +\\
    \mathrm{{\alpha _{prd}}{\mathcal{L}_{prd}} + {\alpha _{vel}}{\mathcal{L}_{vel}}}
\end{multline}
where $\mathrm{\mathcal{L}_{NNM}}$ is the overall loss function for NNMs-NF-DNN framework  and each of the weights of the loss function is presented in Table.~\ref{tab:1}. $\mathrm{\mathcal{L}_{\mathcal{N}}}$, $\mathrm{\mathcal{L}_{corr}}$, $\mathrm{\mathcal{L}_{evol}}$, $\mathrm{\mathcal{L}_{prd}}$, and $\mathrm{\mathcal{L}_{vel}}$ are loss functions corresponding to reconstruction in original coordinates, independence between modal coordinates, evolution~(dynamics) in latent space, prediction in original coordinates, and state-space format in latent space respectively which have been expressed in detail as below: 
\begin{enumerate}
\itemsep=3pt
\item Normalizing Flows loss function: negative log-likelihood (NLL). Normalizing Flows probability density estimation is the first loss function. Generally, the mean squared loss function is a log-likelihood loss function. Therefore, NLL is sufficient for reconstruction: 
\begin{equation}
   {\mathcal{L}_{\mathcal{N}}=-\frac{1}{D}\sum_{X\in D} log\hspace{0.1cm} \mathcal{P}(X)} 
\end{equation}
 where $\textit{X}$ is the original state spaces and ${D}$ denotes the training dataset. 
\item Although Normalizing Flows alone is able to decompose a coupled vibration, we still need auxiliary loss functions in order to enhance the decomposition of modal coordinates. In order to make NNM modal coordinates independent, modal-uncorrelated loss functions are presented as follows~(${\mathcal{L}_{corr}}$):
\begin{equation}
\begin{split}
\mathrm{\frac{1}{ns}\sum_{i=1}^{i=ns}||Corr\left(\bold{p} \right),I_{n\times n}||^{(i)}_{{\text{MSE}}}}  \\
\mathrm{\frac{1}{ns}\sum_{i=1}^{i=ns}||Corr\left(\bold q\right),I_{n\times n}||^{(i)}_{{\text{MSE}}}}  \\
\mathrm{\frac{1}{ns}\sum_{i=1}^{i=ns}||Corr\left(\bold {\dot{p}}\right),Corr\left(\bold p\right)||^{(i)}_{{\text{MSE}}}}
\end{split}
\end{equation}
where $\mathrm{I}$ and $\mathrm{Corr}$ are identity matrix and correlation matrix, respectively; $\mathrm{p}$ is the displacement matrix :~${ \mathrm{p}=[p_{1},p_{2},...,p_{n}]}$ and $\mathrm{q}$ is the velocity matrix :~${ \mathrm{q}=[q_{1},q_{2},...,q_{n}]}$. Each ${p_{i}}$ or ${q_{i}}$ is a vector of length $\mathrm{T}$. ${ \dot{p}}$ is ${\frac{\Delta \boldsymbol p}{\Delta t}}$ (time derivative, ${\Delta t}$ is also given to the network as an input information) and $\mathrm{s}$ is the number of degrees of freedom of the system. It should be noted that, $\mathrm{||~,~||_{MSE}}$ denotes the mean squared error between two matrices or vectors: for example between the reconstructed trajectory and original trajectory and $\mathrm{ns}$ is the number of training samples and $\mathrm{(i)}$ refers to the sample index number.
\item Dynamics block to identify the evolution function: Evolution in latent subspace (nonlinear dynamics). In dynamics block (grey color in Fig.~\ref{FIG:3}), the networks use the initial time response of each example of training to predicate the evolution of system states recursively. It can be implemented by minimizing the residual of the expression below:
\begin{equation}
 \mathrm{\mathcal{L}_{evol}}=\frac{1}{ns}\sum_{i=1}^{i=ns}||\mathcal{N}^{-1}\left(\textit {X}^{t+1}\right),  G\left(\mathcal{N}^{-1}\left(\textit X^{t}\right)\right)||^{(i)}_{{\text{MSE}}}   
\end{equation}
 where ${G}$ is the dynamics block which can be modelled as a nonlinear embedded dynamics with nonlinear activation functions (Relu function). We minimize the loss of ${m}$ time-step prediction:
 \begin{equation}
     {\frac{1}{ns}\sum_{i=1}^{i=ns}||\mathcal{N}^{-1}\left(\textit X^{t+m}\right), G\left( G\left(G...\left(\mathcal{N}^{-1}\left(\textit X^{t}\right)\right)\right)\right)||^{(i)}_{{\text{MSE}}}}
 \end{equation}
  where the state space has to pass ${m}$ times through the nonlinear dynamics block (${G}$).
\item Prediction by incorporating NF and dynamics block: Prediction in original coordinates. After evolution in latent coordinates, NF should transform it back to the original coordinates by minimizing: 
\begin{equation}
 \mathrm{{\mathcal{L}_{prd}}=\frac{1}{ns}\sum_{i=1}^{i=ns}||\textit X^{t+1}, \mathcal{N}\left(G\left(\mathcal{N}^{-1}\left(\textit X^{t}\right)\right)\right)||^{(i)}_{{\text{MSE}}}}   
\end{equation}
 or generally for ${m}$ time-step prediction, we minimize ${\frac{1}{ns}\sum_{i=1}^{i=ns}||\textit X^{t+m}, \mathcal{N}\left(G\left(G\left(G...\left(\mathcal{N}^{-1}\left(\textit X^{t}\right)\right)\right)\right)\right)||^{(i)}_{{\text{MSE}}}}$
\item Velocity loss.  As mentioned before, each pair of latent dimensions should be displacement and velocity fields of a modal coordinate~(${p,q}$). To enforce NNMs-NF-DNN to learn under this constraint, a corresponding loss function is integrated:\begin{equation}
\mathrm{{\mathcal{L}_{vel}}=\frac{1}{ns}\sum_{i=1}^{i=ns}||\frac{\Delta{\textit{p}_i}}{\Delta t}, \textit{q}_i||^{(i)}_{{\text{MSE}}}}
\end{equation}
\end{enumerate}
\begin{table}
	\caption{Weights of loss functions.}
	\label{tab:1}       
	\begin{tabular}{lp{1cm}p{1cm}p{1cm}p{1cm}}
		\hline\noalign{\smallskip}
		$\alpha_{\mathcal{N}}$  & $\alpha_{\mathrm{evol}}$ & $\alpha_{\mathrm{prd}}$  & $\alpha_{\mathrm{corr}}$ & $\alpha_{\mathrm{vel}}$  \\
		\noalign{\smallskip}\hline\noalign{\smallskip}
		1 & 1000 & 1000 & 1 & 1  \\

		\noalign{\smallskip}\hline
	\end{tabular}
\end{table}
\begin{figure}[htbp]
	\centering
	\includegraphics[width=0.5\textwidth]{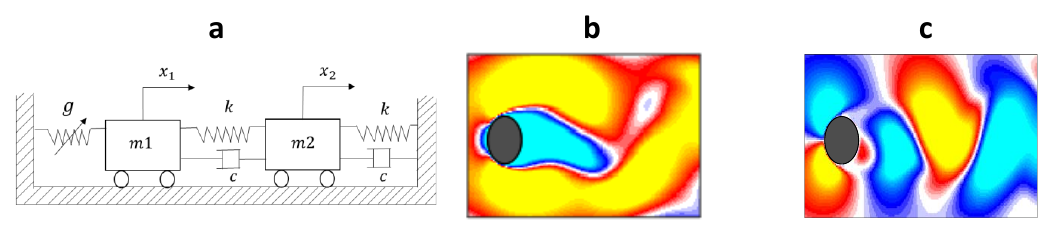}
	\caption{Case studies: \textbf{a}: 2 DOF Duffing system \textbf{b}: Stream-wise velocity over a cylinder \textbf{c}: Transverse velocity over a cylinder }
	\label{case_studies}
\end{figure}

\begin{figure*}[htbp]
	\centering
	\includegraphics[width=0.9\textwidth]{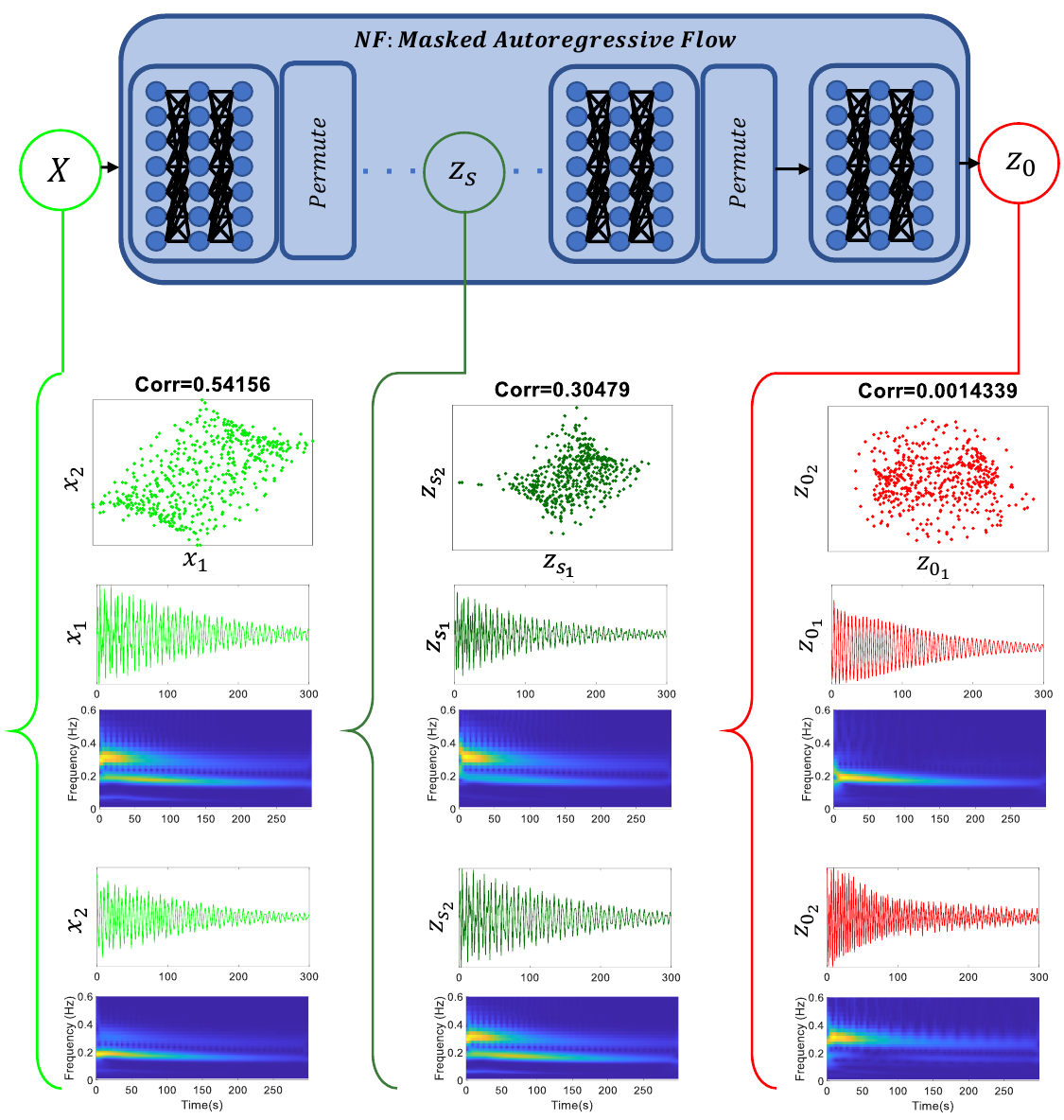}
	\caption{Mode decomposition with NF model: We input the original coordinates of a 2-DOF Duffing system ($x_1$ and $x_2$ of $X$) and then obtain the decomposed modal coordinates at the last layer of NF~($Z_0$). The dependency decreases over the NF layers as the outputs after layer $s$ ($Z_s$) have less correlation magnitudes compared to the original coordinates correlation value. \textcolor{black}{The wavelet plot of each coordinate indicates whether it exhibits single-mode oscillation or a combination of multiple modes}. Note that the nonlinearity is also observable when the frequency changes over time in the wavelet plots.}
	\label{FIG:5}
\end{figure*}


\subsubsection{Flow passing a cylinder}
A limitation of Normalizing Flows is one-to-one mapping~(Fig.~\ref{fig-one-to-one}) which makes high-dimensional systems such as flow fields difficult to model by this algorithm. To address this challenge, we first apply POD to the original flow fields to have a reduced spatial representation for decreasing computational cost. Since POD modal coordinates are linearly independant, there remain nonlinear dependency. As depicted in Fig.~\ref{FIG:9} and Fig.~\ref{FIG:10} , two POD modal coordinates of a velocity field of flow are linearly independent while the nonlinear dependency is obvious as they are bonded in a circle. To overcome this dependency, we input these POD coordinates to Normalizing Flows model and separate them naturally by leveraging the inherent independency of latent coordinates of Normalizing Flows. \par
Because the focus of studying the flow field is nonlinear mode decomposition, the loss functions are limited to decomposition and reconstruction losses as described for the 2-DOF Duffing system. The only difference in decomposition loss function is that since the governing equation is Navier-Stokes equations (a PDE) the latent coordinates are not pairs as Duffing systems (state-space equations). Therefore, to make POD modal coordinates independent, the modal-uncorrelated loss functions are presented as:
\begin{equation}
    {\frac{1}{ns}\sum_{i=1}^{i=ns}||Corr\left(Z_0\right),\mathrm{I}_{n\times n}||^{(i)}_{{\text{MSE}}}}
\end{equation}
where  $\mathrm{I}$,  ${Corr}$, ${Z_{0}}$, and $\mathrm{n}$  are identity matrix, correlation matrix, POD modal coordinates, and the number of modal coordinates, respectively. As stated before, ${ns}$ is the number of training samples.

\subsubsection{Network architecture and training}
In the presented NNMs-NF-DNN for \textit{Duffing systems}, there are two main models: 1 - Normalizing Flows 2 - Dynamics block. Each model performs the following tasks:\par
\textit{Normalizing Flows}: This model converts the original coordinates into modal coordinates (forward \& inverse modal transformation). The output of this model is the latent modal coordinates, and these coordinates will then be passed through the dynamics block. Therefore, the loss functions associated with this model are $\mathrm{\mathcal{L}_{rec}}$, $\mathrm{\mathcal{L}_{prd}}$, $\mathrm{\mathcal{L}_{corr}}$, $\mathrm{\mathcal{L}_{vel}}$.\par
\textit{Dynamics block}: This model represents the dynamics of systems by mapping intrinsic modal coordinates to some specified time steps in advance. This can be achieved by training two loss functions in the overall presented framework: $\mathrm{\mathcal{L}_{evol}}$ and $\mathrm{\mathcal{L}_{prd}}$. Dynamics block model is a multi-layer perception model, with 4 dense layers each containing 256 neurons. It should be noted that, for flow case studies~(Fig.~\ref{FIG:4}), we only have NF model and we apply POD first on the original flow fields as pre-processing dimension reduction phase; after mode decomposition we perform the POD inversion to get back to the original flow fields.

The autoregressive models of Normalizing Flows are powerful models for estimating probability densities. The Normalizing Flows model used in this work consists of several dense layers, each with specified number of neurons which are reported in Table.~\ref{tab:2}. Nonlinear activation functions are used because nonlinear modal transformations are sought. In Tensorflow, there are different nonlinear activation functions such as Relu, Sigmoid, and Tanh. Our networks utilize the Relu function, which has a faster training run time\cite{Lusch2018}. Each autoregressive layer is followed by a permutation layer since Normalizing Flows layers only operate on a portion of the data, whereas the remaining does not change when passing through these layers. Therefore, we permute the data spaces so that all data are subjected to nonlinear transformations through the network (Fig.\ref{FIG:3}).

The Adam optimizer with a slow learning rate of $\alpha=1e-5$ is used for both models~(dynamics model and Normalizing Flows model). The Xavier initialization method~\cite{glorot2010understanding} is used to initialize the weights of each model. Hidden layers are in the format of ${{Wa + b}}$ followed by a nonlinear activation function where ${W}$ and ${b}$ are weights and biases respectively and ${a}$ refers to input data. The Xavier initialization method generates a random number that is distributed uniformly along a range of $-\frac{1}{\sqrt{\eta}}$ and $\frac{1}{\sqrt{\eta}}$, where $\mathrm{\eta}$ refers to the number of inputs to the node. We analyze the performance of DNN across a variety of training sessions~(hyperparameters-tuning).  It has been examined different sets of hyperparameters (weights of loss functions) and the results are based on the hyper-parameters associated with minimum testing errors (Table.~\ref{tab:1}). 
\begin{figure*}[htbp]
	\centering
	\includegraphics[width=1.1\textwidth]{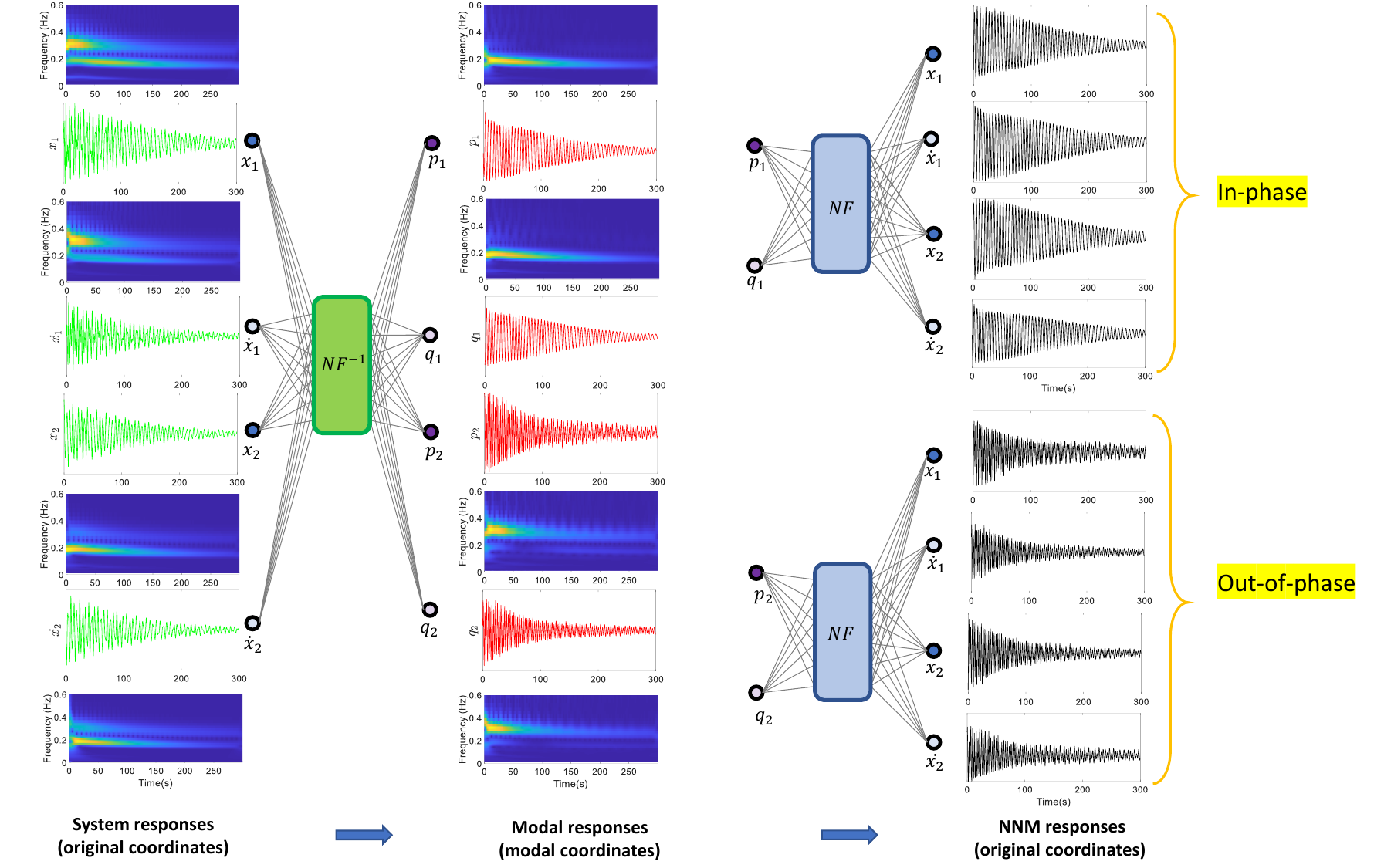}
	\caption{An illustration of mode decomposition from damped nonlinear system response using Normalizing Flows DNN. First, the inverse of NF transforms the input system response ${\mathrm{x}} = {\left[ {{x_1},{\dot x_1},{x_2},{\dot x_2}} \right]}$ to modal space where each pair of modal displacement ${p_i}$ and modal velocity ${q_i}$ has a distinct frequency. Second, each pair of modal response (${p_i}$ and ${q_i}$) are transferred back to original coordinates separately using NF model which ultimately outputs the corresponding modal coordinates in the original space~(in-phase and out-of-phase modal coordinates).}
	\label{FIG:6}
\end{figure*}
\begin{figure*}[htbp]
	\centering
	\includegraphics[width=0.9\textwidth]{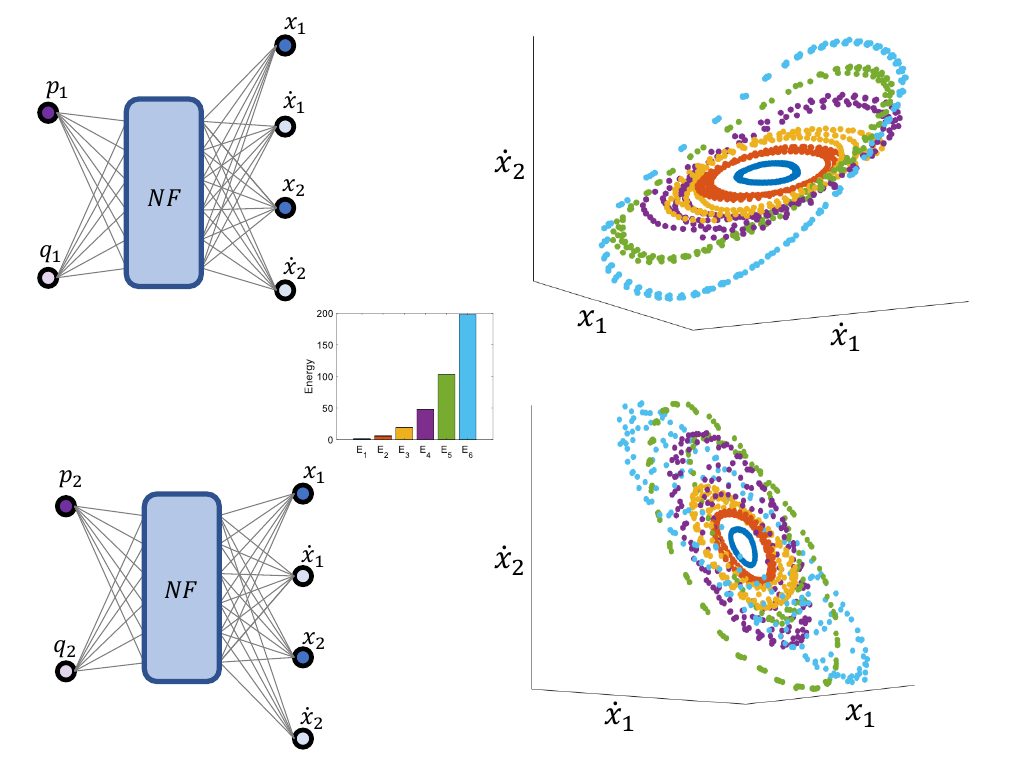}
	\caption{ The NNMs' invariant manifolds of a conservative 2-DOF Duffing system identified from response data only using the physics-integrated NF-DNN framework. The in-phase~(top manifold plot) and out-of-phase (bottom manifold plot) manifolds are obtained by transferring back the each pair of latent space separately to the original space.  }
	\label{FIG:7}
\end{figure*}

\begin{figure*}[htbp]
	\centering
	\includegraphics[width=1\textwidth]{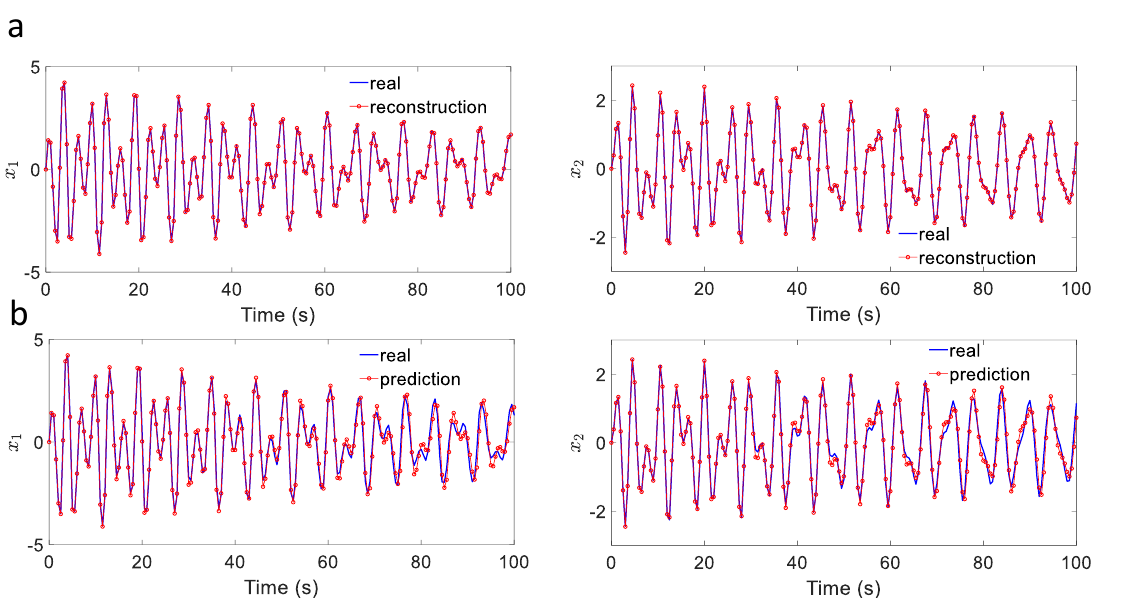}
	\caption{Reconstruction and prediction performance of our presented NF-DNN for a 2-DOF damped Duffing system. \textbf{a} Reconstruction. \textbf{b} Prediction  }
	\label{FIG:8}
\end{figure*}

\begin{table*}
	\caption{Network architecture}
	\label{tab:2}       
	\begin{tabular}{lp{1.5cm}p{2cm}p{2cm}p{1.5cm}}
		\hline\noalign{\smallskip}
		Item &No. of NF layers & No. of Dense layers in each NF layer & No. of neurons in each Dense layer \\
		\noalign{\smallskip}\hline\noalign{\smallskip}
		Duffing system& 6 &3 & 256 &  \\
		Transverse velocity field& 8 &3 & 512 &  \\
		Streamwise velocity field& 8 &3 & 512 &  \\
		
		\noalign{\smallskip}\hline
	\end{tabular}
\end{table*}

\section{Result and Discussion}
\subsection{2-DOF Duffing oscillator}
Duffing systems (see Fig. \ref{case_studies}) are widely used in research on dynamic analysis~\cite{lin2004non,nijmeijer1995lyapunov}. As one of the case studies, we study a 2-DOF Duffing system with a governing equation as follows:
\begin{equation}\label{eq:14}
\begin{gathered}
{{{\ddot x}_1} + 0.03{{\dot x}_1} + \left( {2{x_1} - {x_2}} \right) + 0.5x_1^3 = 0 \hfill} \\
{{{\ddot x}_2} + 0.01{{\dot x}_1} + \left( {2{x_2} - {x_1}} \right) = 0 \hfill} \\ 
\end{gathered}.
\end{equation}
\par
We first examine the performance of our presentd NNMs-NF-DNN with a 2-DOF nonlinear Duffing system. The state space is input to the network, and Normalizing Flows estimates the PDF of these trajectories by converting a Gaussian distribution as the base distribution to the more complicated distribution (Fig. \ref{FIG:1}). Since each dimension is sampled from a Gaussian distribution, therefore each dimension is independent. The unique feature of the Normalizing Flows enable to decompose a coupled vibration into independent components in a natural manner throughout the PDF estimation process. However, since velocity is merely a time derivative of displacement, the latent dimension ought to follow the same relationship. To restrict the network to have this feature, we use velocity loss function, but since this loss affects the decomposition of modal coordinates, we use correlation loss function to balance the network so that it has both decomposition and state-space presentation in latent space as NNM coordinates.\par
Fig. \ref{FIG:5} (a) illustrates the process of modal decomposition throughout the layers of Normalizing Flows. The wavelet graphs show that in the original coordinates (light green scatter plot), there is a coupled vibration with two modal frequencies. We have quantified the dependency of displacement vectors using the correlation coefficient ($\mathrm{Corr=0.541}$). Moving through the Normalizing Flows layers, it is observed that the dependency decreases as a good level of decomposition is achieved in the last layer (red scatter plot). \par
Fig. \ref{FIG:6} illustrates the process of single-mode reconstruction using Normalizing Flows. The original coordinates are first converted to modal coordinates as ${p_i}$ and ${q_i}$. In this case, two independent modal coordinates exist for this 2-DOF Duffing system. To illustrate each modal coordinate in the original coordinate system, we only focus on the corresponding modal coordinate in the latent space while freezing the other pair of modal coordinates. Subsequently, we utilize the direct inverse feature of NF to attain the corresponding original modal coordinates from latent space. The in-phase and out-of-phase modal coordinates are illustrated in Fig. \ref{FIG:6} for a 2-DOF system with system parameters: $\mathrm{m_1=m_2=1}$, $\mathrm{k=1}$, $\mathrm{c=0.001}$, and  $\mathrm{g=0.03}$~(Fig. \ref{case_studies}).\par
An illustration of the identified NNMs for an undamped 2-DOF Duffing system based on different levels of energy is presented in Fig.~\ref{FIG:7}. The NNM manifolds of each energy level correspond to a specific initial condition and their corresponding NNMs-NF-DNN identify their respective NNM manifolds. A single-mode reconstruction is performed by using the corresponding pair of latent coordinates, which represent modal coordinates to identify in-phase and out-of-phase manifolds. Increasing the energy of the system results in higher nonlinearity, where the in-phase and out-of-phase mode shapes undergo a change from flat (planar) to curved manifolds (Fig.~\ref{FIG:7}), which is in agreement with analytical results~\cite{peeters2011modal}.\par
Fig.~\ref{FIG:8} (a) illustrates the reconstruction ability of the presented approach where the reconstruction is in excellent agreement with the true response. Since Normalizing Flows is an invertable network, a decoder model like autoencoders is not required. This is a key feature of NF to allow for both encoder and decoder in a single model. Therefore, fewer parameters are needed to train the model. The decoder in autoencoder DNNs is only an estimation of an inverse encoder, so there are errors when decoding the latent space to the original space; whereas in Normalizing Flows, there are bijective layers which guarantees the direct mathematical inverse of forward transformation. Fig.~\ref{FIG:8} (b) illustrates the prediction ability of the NNMs-NF-DNN, showing excellent prediction accuracy. There are 500 time steps for prediction in this range. The framework receives the initial state and predicts the future 499 time steps recursively. Since each step is the prediction based on the estimated previous time step, it requires a precise prediction at each time step to avoid error accumulation over time.
\subsection{Flow fields passing over a cylinder}
We study flow fields as another case study. Specifically, we consider a two-dimensional flow field over a cylinder, which is a typical example used in many existing works, \cite{raissi2020hidden,murata2020nonlinear}, to validate the feasibility of the presented method. 
\par
In this case study, flow passes through the cylinder and creates some vortex shading in its wake, which is known as a Karman vortex street. It is a steady-state flow in which the Reynolds number varies between ${Re_{D}=100}$ and ${Re_{D}=200}$. The governing equation is the Navier-Stokes equation (NS):

\begin{equation} 
\begin{split}
&\mathrm{\nabla\cdot U=0} \\
&\mathrm{\frac{\partial U}{\partial t}=-\nabla\cdot (UU)-\nabla p+\frac{1}{R_{D}}\nabla^{2}U}
\end{split}
\end{equation}
where $\mathrm{U}$ and $\mathrm{p}$  are velocity and pressure, respectively. Stream-wise and transverse velocity are assigned as $\mathrm{u}$ and $\mathrm{v}$ correspondingly. No-slip boundary condition is applied. The channel has a grid size of 96 by 192. \par

Traditionally, POD is a linear technique widely used to analyze fluid flows in fluid dynamics~\cite{holmes2012turbulence} . Inputs to this algorithm include snapshots of flow properties ($\mathcal{M}(\zeta,t)$) such as temperature, pressure, velocity, etc. The output is a set of orthogonal modes representing the dominant spatial characteristics of the flow. The formulation is as follows: 
\begin{equation}
    \mathrm{\mathcal{M}(\zeta,t)- \bar {\mathcal{M}}(\zeta)=\sum_{j} a_{j}(t)\phi_{j}(\zeta)}
\end{equation}\label{eq:2}
where $\mathrm{\bar {\mathcal{M}}(\zeta)}$ is the temporal mean of flow field, $\mathrm{\phi_{j}(\zeta)}$ and $\mathrm{a_{j}}$ are modes and expansion coefficients, respectively \cite{taira2017modal}. Flow fields are reconstructed with superposition of a few dominant POD modes when the fluid system is not highly nonlinear, but the resulting mode lacks dynamics information about fluid flow since POD is merely a spatial transformation that captures spatial patterns in the original flow fields.

We test the nonlinear mode decomposition capabilities of the presented Normalizing Flows deep neural network~(NF-DNN) for the flow field. As discussed, because NF is one-to-one mapping~(Fig.~\ref{fig-one-to-one}), direct application of Normalizing Flows to the high-dimensional flow field would be computationally expensive and time-consuming. Therefore, we utilize POD in pre-processing to reduce the spatial dimensions while retaining the most important features of flow. In the pre-processing phase, only 10 POD modal coordinates are retained. 

\noindent\textbf{Remarks:} Even though the correlation coefficient, which is a linear metric of dependency, is zero for the first two POD modal coordinates, the mutual information value indicates that they are dependent in a \textit{nonlinear} manner.~Briefly, assuming $\mathrm{(A,B)}$ are two random variables with values over $\mathrm{A\times B}$, if their joint distributions is $\mathrm{P_{AB}}$ and the marginal distributions are $\mathrm{P_{A}}$ and $\mathrm{P_{B}}$, mutual information may be defined as: 
\begin{equation}
\mathrm{MI(A;B)=D_{KL}(P_{(A,B)}||P_{A} \otimes P_{B})}
\end{equation}
where $\mathrm{D_{KL}}$ is Kullback–Leibler divergence~\cite{csiszar1989geometric}. Unlike Pearson correlation coefficients which only detect linear relationships between variables, mutual information functions can detect nonlinear relationships between variables. Therefore, the focus here is to further make the POD modal coordinates independent by leveraging the key feature of NF which provides exact independent coordinates in the latent space. \par
\begin{figure*}[htbp]
	\centering
	\includegraphics[width=0.95\textwidth]{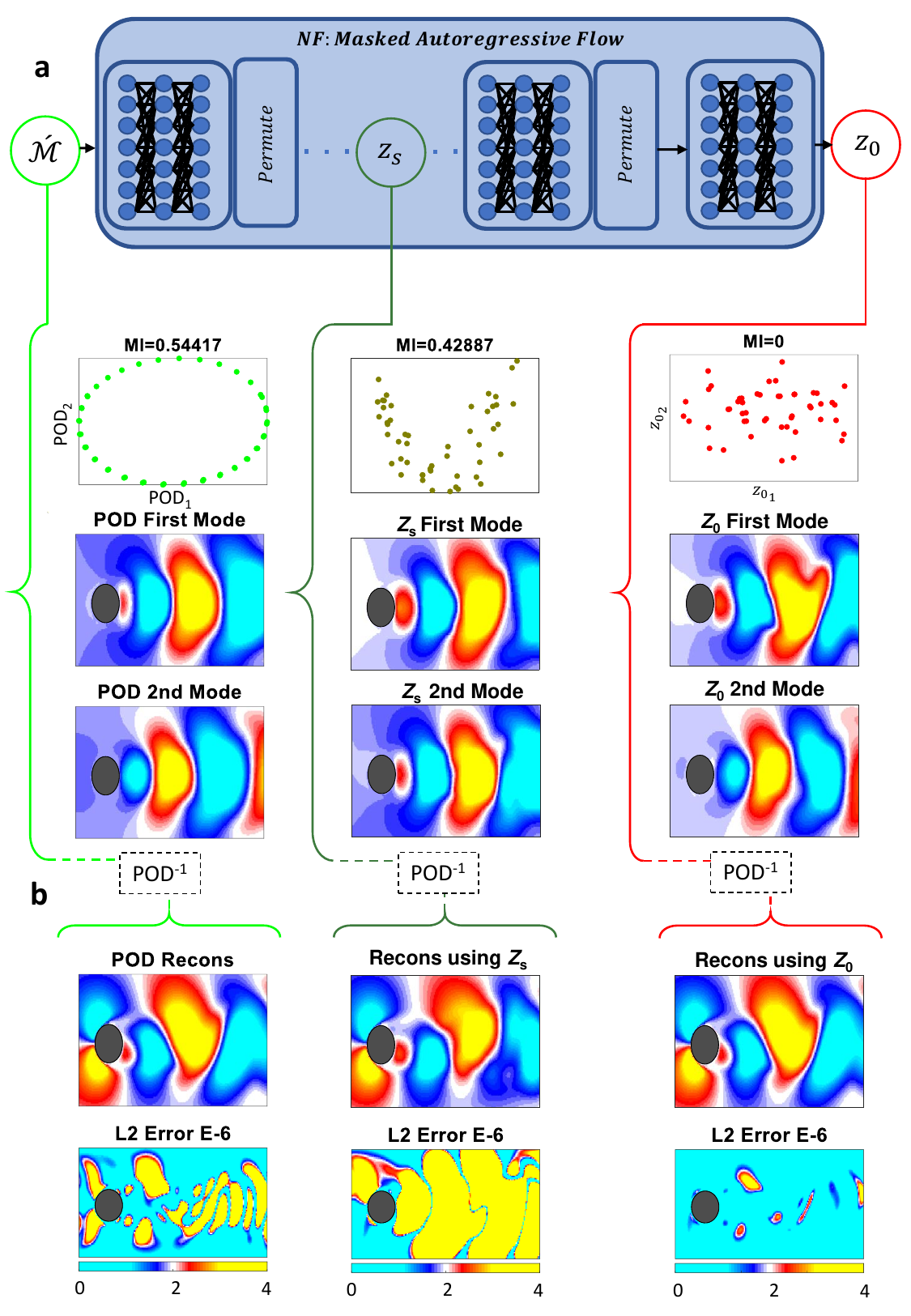}
	\caption{\textbf{(a)} Nonlinear mode decomposition with NF model for \textbf{transverse} flow field: The POD modal coordinates are input to the model ($\mathrm{POD_1}$ and $\mathrm{POD_2}$ which are equal to $X$ and then the decomposed modal coordinates are obtained at the last layer of NF~($Z_0$)). The dependency decreases over the NF layers as the outputs after layer $s$ ($Z_s$) have a smaller correlation magnitude compared to the original POD modal coordinates. The first and second spatial modes for each space are also shown. The output spatial modes~(NF modes) are nonlinear versions of POD modes as the spatial patterns are twisted compared to POD modes. \textbf{(b)} Reconstruction performance of original POD modes, NF modes, and a space between original and final layer ~($Z_s$) with their reconstruction errors.}
	\label{FIG:9}
\end{figure*}
\begin{figure*}[htbp]
	\centering
	\includegraphics[width=0.95\textwidth]{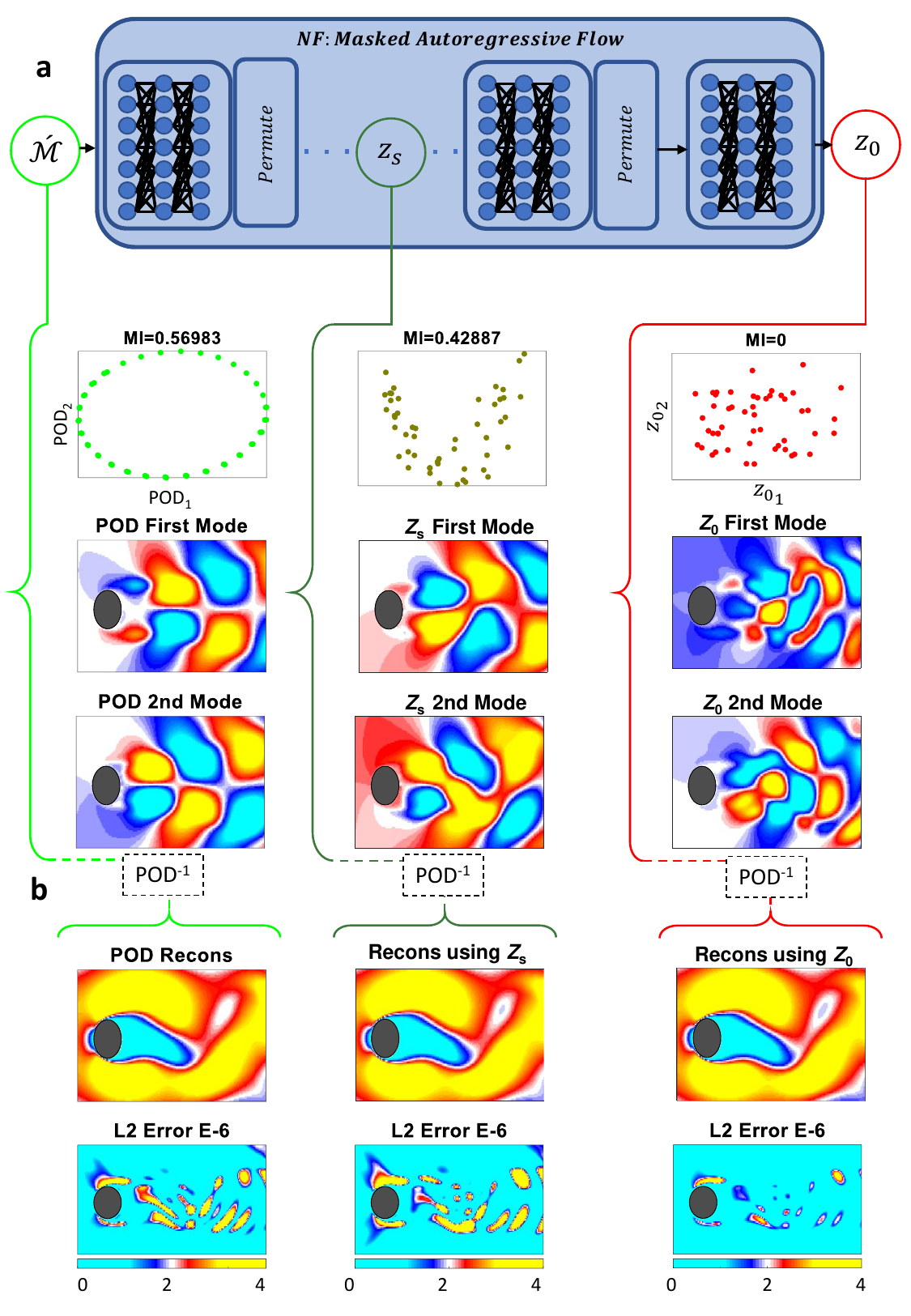}
	\caption{\textbf{(a)} Nonlinear mode decomposition with NF model for streamwise flow field: The POD modal coordinates are input to the model ($\mathrm{POD_1}$ and $\mathrm{POD_2}$ which are equal to $X$ and then the decomposed modal coordinates are obtained at the last layer of NF~($Z_0$)). The dependency decreases over the NF layers as the outputs after layer $s$ ($Z_s$) have a smaller correlation magnitude compared to the original POD modal coordinates. The first and second spatial modes for each space are also shown. The output modes~(NF modes) are nonlinear versions of POD modes as the spatial patterns are twisted compared to POD modes. \textbf{(b)} Reconstruction performance of original POD modes, NF modes, and a space between original and final layer ~($Z_s$) with their reconstruction errors.}
	\label{FIG:10}
\end{figure*}

A latent space has been decomposed after the network has been trained using $10$ POD modal coordinates. We now use two latent coordinates and deactivate the remaining coordinates by setting their values to zero and then reconstruct the original data using both NF and POD. Based on the same number of independent modal coordinates, i.e., $2$, the reconstruction ability of POD and Normalizing Flows is compared in Fig.~\ref{FIG:9}(b) and Fig. 
 \ref{FIG:10}(b) for streamwise and transverse velocity, respectively. It is evident that NF achieves much higher reconstruction accuracy than POD (evaluated by the $L2$ errors), indicating that the decomposed NF coordinates contain much more nonlinear dynamics features about the flow field. 
 
 \textcolor{black}{The reconstruction accuracy of POD and NNMs for both streamwise and transverse velocity is quantified in Table. \ref{tab:3}}. Additionally, each of Fig.~\ref{FIG:9} and Fig.~\ref{FIG:10} shows the first and second spatial modes of POD, modes obtained in the middle layer of NF, and modes related to the final layer which represents the nonlinear version of POD modes obtained by Normalizing Flows, respectively. NF mode shapes depict the nonlinear version of POD modes, which contains the features of the first 10 POD modal coordinates and is captured and represented in only two Normalizing Flows modal coordinates. Fig.~\ref{FIG:9}(a) and Fig.~\ref{FIG:10}(a) illustrate how mutual information between POD modal coordinates decreases as passing through the NF layers from a circle (high dependency) to a random distribution (low dependency).

 \begin{table*}
	\caption{MSE of reconstruction}
	\label{tab:3}       
	\begin{tabular}{lp{1.5cm}p{2cm}p{2cm}p{1.5cm}}
		\hline\noalign{\smallskip}
		Item &POD &NNMs  \\
		\noalign{\smallskip}\hline\noalign{\smallskip}
		Streamwise velocity& 7.1 e-3& 2.9 e-6  \\
		Transverse velocity field& 2.5 e-4& 3.2 e-6   \\

		\noalign{\smallskip}\hline
	\end{tabular}
\end{table*}


\section{Conclusion}
In this study, we employed the unique features of Normalizing Flows~(NF) approach, which allows it to learn the intricate underlying distribution of complex data from a simpler independent distribution. We aimed to utilize this method as a nonlinear modal analysis technique for representing nonlinear normal modes (NNMs). To achieve this, we conducted multiple case studies involving various nonlinearities, including the Duffing system and fluid flows. As we advance through the NF layers, our observations indicate a decrease in the dependency of the original coordinates. This enables us to achieve fully decomposed modal coordinates in the latent space, effectively representing the NNMs with the aid of other loss functions.  To assess the effectiveness of our approach for fluid flows, we compared the results of a flow over a cylinder to those obtained from  Proper Orthogonal Decomposition (POD) as a linear modal analysis technique. We found that the nonlinear version of POD, acquired through the utilization of NF, contains more substantial information about a flow field. As a result, the reconstructions obtained through this approach exhibit greater accuracy compared to those achieved through POD reconstruction. Additionally, we demonstrated the model's capability for trajectory predictions with the aid of embedded dynamics block, as exemplified with Duffing systems.


It is essential to note that our framework may have some limitations. Firstly, the training of the NF model is not entirely stable, requiring the use of a small learning rate to mitigate this limitation. Consequently, a substantial number of epochs need to be considered, leading to a relatively time-consuming training phase. Additionally, the one-to-one mapping characteristic of NF makes it computationally demanding when applied to high-dimensional dynamical systems, such as fluid flows. Another significant challenge is the limited range of nonlinearity or energy levels present in our datasets. To overcome this limitation and accurately model higher nonlinear systems, it is essential to include a more extensive and diverse range of nonlinearity and energy levels in future data collection. Lastly, in our current study, we assumed the absence of internal resonance in the studied dynamical systems . To address potential scenarios with internal resonance in future research, specific modifications to the architecture of presented DNNs will be required to accommodate and properly capture the effects of internal resonance. Also, The intersection of smart system design principles and nonlinear dynamics modeling aligns with innovative strategies for sustainability, particularly in adapting dynamic solutions for environmental and structural applications \cite{shafa6ranking,shafa6smart}.
\section*{Data Availability}
The data that support the findings of this study are available from the corresponding author upon reasonable request.
\section*{Funding}
This research was partially funded by the Physics of Artificial Intelligence Program of U.S. Defense Advanced Research Projects Agency (DARPA), the Michigan Technological University faculty startup fund, and the Michigan Tech Excellence Fund.

%
\section*{Conflict of interest}
The authors declare no conflict of interest.

\bibliographystyle{unsrt}

\bibliography{my-collection}   

\end{document}